\documentclass[aps,prl,reprint,amssymb,groupedaddress,twocolumn]{revtex4}
\usepackage{graphicx} 
\usepackage{xcolor}
\usepackage{subfigure}
\usepackage{hyperref,hypcap}
\usepackage{braket}
\usepackage{amsmath}
\usepackage{multirow}
\usepackage{bm}
\usepackage{hhline}
\newcommand{\CT}{$\mathcal{C}_2\mathcal{T}$ }
\newcommand{\CTs}{$\mathcal{C}_2\mathcal{T}$}

\usepackage[normalem]{ulem}
\newcommand\redout{\bgroup\markoverwith
	{\textcolor{red}{\rule[.5ex]{2pt}{1pt}}}\ULon}

\begin{document}
\title{On the Nature of the Correlated Insulator States in Twisted Bilayer Graphene}
\author{Ming Xie}
\affiliation{Department of Physics, The University of Texas at Austin, Austin, TX 78712, USA}
\author{A. H. MacDonald}
\affiliation{Department of Physics, The University of Texas at Austin, Austin, TX 78712, USA}

\date{\today}

\begin{abstract}
We use self-consistent Hartree-Fock calculations performed in the full 
$\pi$-band Hilbert space to assess the nature of the recently 
discovered correlated insulator states in magic-angle twisted bilayer graphene (TBG).
We find that gaps between the 
flat conduction and valence bands open at neutrality over a wide range of twist angles,
sometimes without breaking the system's valley projected \CT
symmetry.  Broken spin/valley flavor symmetries then enable
gapped states to form not only at neutrality, but also at total moir\'e band filling 
$n = \pm p/4$ with integer $p = 1, 2, 3$, when the twist angle is close to the magic 
value at which the flat bands are most narrow. 
Because the magic-angle flat band quasiparticles are isolated from remote band quasiparticles only for 
effective dielectric constants larger than $ \sim 20$, the gapped states do not 
necessarily break \CT symmetry and as a consequence the insulating states at 
$n = \pm 1/4$ and $n = \pm 3/4$ need not 
exhibit a quantized anomalous Hall effect. 
\end{abstract}

\maketitle

{\em Introduction.}---
A small relative twist between adjacent graphene layers produces a triangular lattice moir\'e pattern
with a spatial periodicity that is inversely related to twist angle.  It was noticed \cite{MorellFlatBand,BMModel}
some years ago that at a series of magic twist angles $\theta$, the moir{\'e} pattern yields
very flat low-energy bands that promise strong electronic correlations.  This promise has now been
realized thanks to recent experimental studies 
\cite{Tutuc2017,CaoInsulator,CaoSuper,YoungDean, Efetov, Gordon, Andrei, Pasupathy, Yazdani}
of bilayers with accurately controlled twist angles that exhibit  
interaction-induced insulating ground states at moir{\'e} band \cite{BMModel} filling factors
$n= \pm p/4$, where $p = -3, \ldots, 3$ is the total charge per moir\'e unit cell.  The insulating 
states are flanked by superconducting domes \cite{CaoSuper, Efetov}.
This exciting discovery has inspired a flurry of theoretical work 
\cite{FuModel, VishwanathMarch, Scalettar, Phillips, SpinLiquid, WangModel, MottAF, Yang, PALee, 
	      Vafek, Rademaker, KoshinoFu, Bascones, Ochi, FuJune, AFMonHoneyComb, 
	       XuBalents, FuSuper, Juricic, Das, Super1, Super2, Super3, Super4}
directed toward achieving a more complete understanding 
of the insulating states and their superconducting satellites.  Previous work on the 
insulating states has been based mainly on an indirect approach that starts by 
identifying effective lattice models for the flat moir\'e bands, and then 
combines these with generalized Hubbard models to address interaction phenomena.  
In this Letter we explore a different approach.

At small twist angles the electronic structure of twisted bilayer graphene can be accurately
described using a continuum model \cite{BMModel, SantosEarly} in which single-particle electronic states 
with a four-level spin/valley internal flavor degree-of-freedom
are approximated by four-component envelope function spinors that specify $\pi$-orbital
amplitudes on the bilayer's four sublattices.
The simplest version of the continuum model \cite{BMModel}
adds a spatially periodic interlayer 
hopping term to isolated layer $\pi$-orbital Dirac models.
This moir\'e band Hamiltonian is spin-independent, and its projections onto graphene's two valleys
are related by time-reversal symmetry.  Up to an overall energy scale, its spectrum depends 
on a single twist-angle dependent parameter 
$\alpha = w/\hbar vk_{\theta}$ where $w \approx 110$ meV is an inter-layer tunneling 
amplitude, $v \approx 10^{6} $ m/s
is the Dirac velocity,
$k_{\theta} = 2 K \sin(\theta/2)$ is the momentum 
separation between the Brillouin-zone (BZ) corners in different layers, 
and $K$ is the single-layer BZ corner momentum magnitude.

\begin{figure}[h]
	\begin{center}
		\includegraphics[width=\columnwidth]{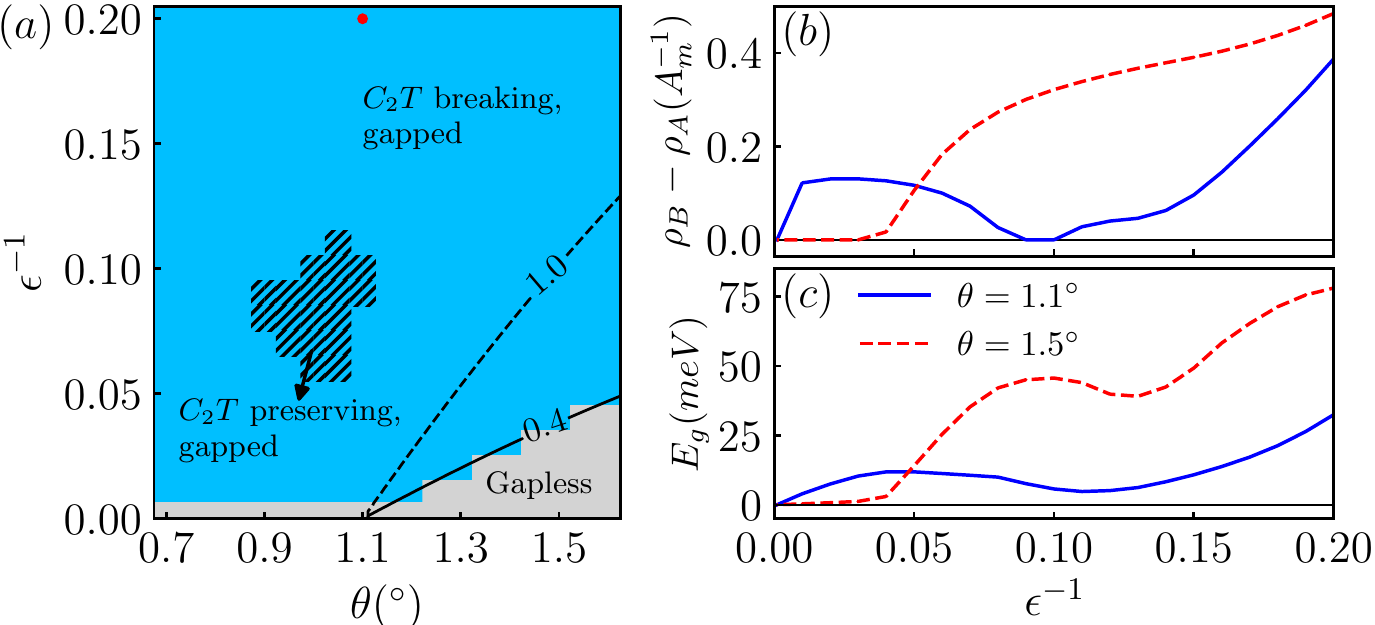}
		\caption{
		Phase diagram of neutral bilayers as a function of Coulomb interaction strength, 
		                characterized by inverse dielectric constant $\epsilon^{-1}$, and twist angle $\theta$. 
		                (a)  Insulating states (blue regions) appear for ever weaker interactions as the 
                               narrow-band magic angle regime near $\theta \sim 1.1^{\circ}$
                                is approached.  Above the magic angle, the bilayer's Dirac points are gapped when the 
                               effective fine structure constant $\alpha^*=e^2/\epsilon\hbar v_D^*$ exceeds $\sim 0.4$ (solid line),
                               where $v_D^*$ is the Dirac velocity of the twisted bilayer,. 
                                ($v_D^*$ goes to zero as the magic angle is approached \cite{BMModel}.) 
                               For $\epsilon^{-1}  \gtrsim 0.05$ remote band degrees of freedom play a role in determining 
                               the phase diagram details, enabling in particular insulating states that do not break  \CT symmetry (hatched 
                               region of the phase diagram).
                        (b)  \CT-breaking order parameter $\rho_B-\rho_A$ (i.e., sublattice polarization) and 
                        (c) global energy gap as a function of  interaction strength $\epsilon^{-1}$ 
                        for $\theta=1.1^{\circ}$ (blue solid line) and $1.5^{\circ}$ (red dashed line).
                        }
		\label{Summary}
	\end{center}
\end{figure}

The valley-projected moir\'e flat bands in TBG occur
not singly, but in valence/conduction pairs connected by two symmetry-protected  
linear Dirac band crossings.
Importantly the two Dirac points of weakly coupled bilayers carry the same chirality.  
This property implies \cite{VishwanathWannierObstructions,VishwanathMarch, Bernevig} 
that the moir\'e flat bands can be described only by tight binding models with at the very 
least four orbitals per spin/valley flavor per moir\'e unit cell.  
Recent work \cite{VishwanathAugust} suggests that faithful descriptions of 
interaction physics using generalized Hubbard
models may require the inclusion of at  least eight bands per flavor, limiting the motivation for,
approximate lattice models.

In this Letter we report on a study of the correlation induced insulator states in magic angle TBG that starts directly from the 
moir{\' e} band continuum model and accounts for the long-range of the Coulomb interaction between electrons.
Our principle results are summarized in Fig.~\ref{Summary}.
In this figure the largest values of interaction strength parameter $\epsilon^{-1}$ correspond to 
screening by a surrounding hexagonal boron nitride dielectric only.  In practice interactions are sample dependent and 
always weakened by nearby gates.  We find that gapped states occur at
neutrality when the effective fine structure constant $\alpha^*=e^2/\epsilon\hbar v_D^*$ exceeds 
$\sim 0.4$ for twist angle $\theta$ above the magic value, and almost always for twists that are smaller.  
Here $v_D^*$ is the reduced Dirac velocity of the twisted
bilayer which vanishes as the magic angle is approached \cite{BMModel}.
We attribute the smaller value of the critical fine structure constant in twisted bilayer graphene 
than in the corresponding single-layer graphene ($\alpha^* \sim 1$ \cite{Min})
calculations to the non-uniform spatial distribution of flat band orbitals, which enhances interaction effects. 
As in single layer graphene\cite{AllanMonolayer}, these gapped states
break \CT symmetry and have non-zero Berry curvatures, whereas the 
the gapless states preserve \CT symmetry.
For $n= \pm p/4 \ne 0$, gapped states are enabled by broken spin/valley
flavor symmetries and occur over a much narrower range of twist angles.
When the interaction strength is sufficiently strong, gapped states can be opened without breaking
the \CT symmetry that protects band crossings when the Hamiltonian is projected onto the strongly correlated flat bands.
This property is significant because it has implications for the occurrence of quantized anomalous 
Hall effects at band filings $n = \pm 1/4$ and $n = \pm 3/4$. 

{\em Mean field theory.}---\noindent
Our theoretical approach is guided by the experimental \cite{CaoInsulator} 
discovery of insulating states in magic angle TBG that are 
naturally explained by broken symmetries that lift 
the four-fold spin/valley degeneracy of the band Hamiltonian, and do not require broken translational symmetry.  
In most cases the ground states of insulators can be described using 
Hartree-Fock mean-field theory.
The difficulty in the TBG case compared to the familiar case of atomic-scale insulators,
is that the Hamiltonian does not contain strong attractive 
potential terms within each unit cell that select particular high-weight atomic or ionic configurations.
To understand the nature of the insulating states, we must perform unbiased Hartree-Fock
calculations in the full \cite{footnote} $\pi$-orbital Hilbert space, placing no restrictions 
on the flavor or position dependence of the model's four-component envelope function spinors.

\begin{table}[t!]
	\begin{tabular}{cccccccc} 
        \hhline{========}
        \multicolumn{1}{c|}{\multirow{2}{*}{E(meV)}} & \multicolumn{3}{c|}{Intra-layer} & \multicolumn{3}{c|}{Inter-layer} & \multirow{2}{*}{Total} \\ \cline{2-7} 
        \multicolumn{1}{c|}{}	& \multicolumn{1}{c|}{Hopping} & \multicolumn{1}{c|}{Hartree} & \multicolumn{1}{c|}{Fock} &  
                                                     \multicolumn{1}{c|}{Hopping} & \multicolumn{1}{c|}{Hartree} & \multicolumn{1}{c|}{Fock} & \\ \hline
          NI       &  1417  &  0  &  -44  &  -2636  &  0  &  -278  &  -1541  \\ 
		 SCHF &  2372 &  0  & -109  &  -3452 &  0  &   -469 &  -1658  \\            \hhline{========}        
	\end{tabular}
	\caption{Total energy per moir\'e unit cell at neutrality in units of meV calculated in the non-interacting (NI)  
		and self-consistent Hartree-Fock (SCHF) ground states for $\theta=1.1^{\circ}$ and $\epsilon=5$.  
		Gapped insulating states with and without broken \CT compete closely, 
		differing in energy by less than 1 meV per moir\'e cell.  Note that Hartree energies do not play an 
		important role in selecting the broken symmetry state.  Both 
		intralayer and interlayer energies grow with the momentum space cut-off, but the difference in energy 
		between non-interacting and interacting states converges.}
		\label{Table1}
\end{table} 

\begin{figure}[t]
	\begin{center}
		\centering
		\includegraphics[width=0.99\columnwidth]{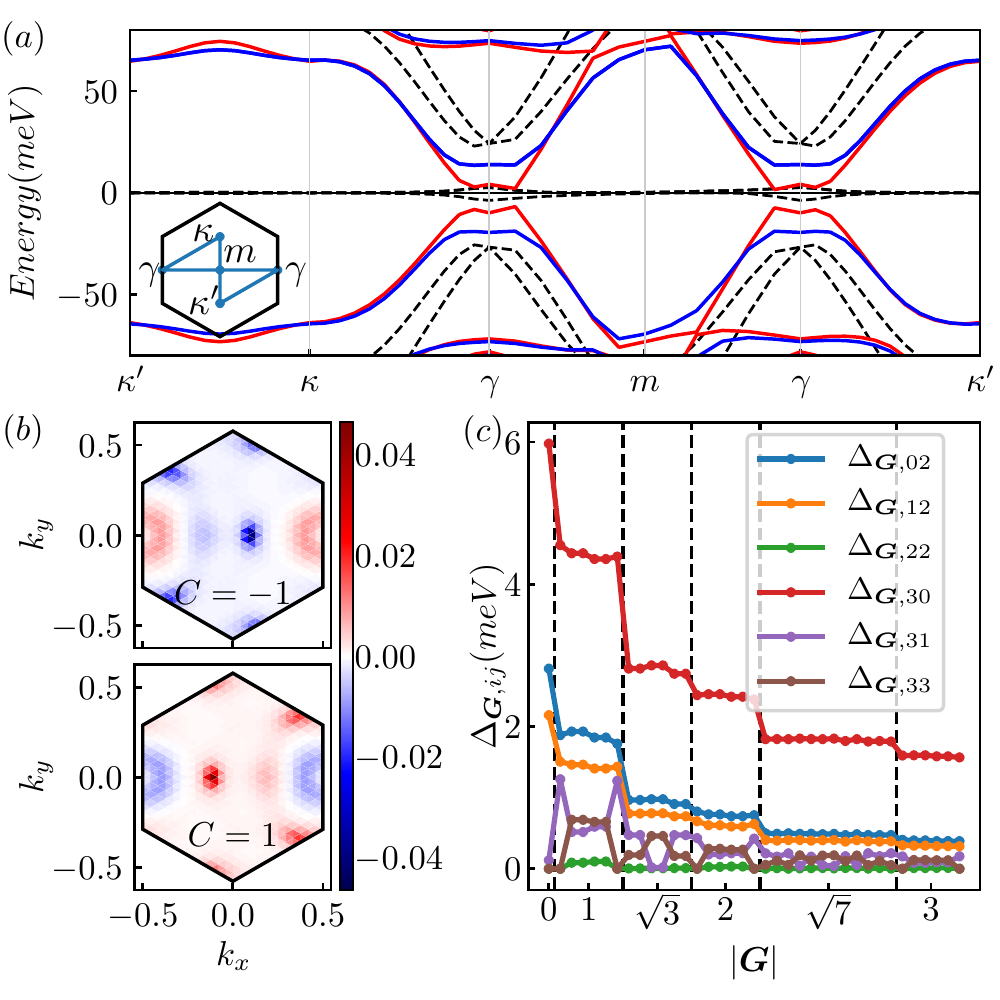}
		\caption{Properties of quasiparticles at $n=0$. Only valley $K$ states are shown. 
			(a) Energy dispersion of the \CT breaking (blue) and \CT preserving (red)
			states at $\theta=1.1^{\circ}$ and $\epsilon=5$.
			The dotted lines illustrate the non-interacting flat bands.  
			The inset specifies the high symmetry lines in the moir\'e BZ (mBZ) along which the band energies have been plotted.
			(b) Berry curvature for lowest conduction band (upper) and the highest valence  band (lower) of the broken 
			${\cal C}_{2}{\cal T}$-symmetry solution.  Features in these plots are related to avoided crossings between 
			the flat conduction band and higher energy remote conduction bands. 
			(c) Order parameters for \CT symmetry breaking as a function of the reciprocal lattice 
			vector $\bm{G}$ magnitude.  The segments marked by vertical dashed lines group 
			$\bm{G}$s with same magnitude.} 
	 \label{Neutrality}
	\end{center}
\end{figure}

A typical self-consistent Hartree-Fock calculation result, for the point  
$\epsilon=5$ and $\theta=1.1^{\circ}$ marked by a red circle in Fig.~\ref{Summary}(a),
is summarized in Fig.~\ref{Neutrality},
which  illustrates quasiparticle dispersion, topology, and \CT breaking order parameters, 
and in Table~\ref{Table1} in which  the ground state energies of  the non-interacting and
interacting cases are partitioned into intralayer and interlayer tunneling and interaction contributions.  
The technical details of these calculations are described in the Supplemental Material. 
All energies are expressed relative to the energy of 
the non-interacting state in the absence of inter-layer tunneling, and a neutralizing background 
charge density is assumed.  
As shown in Fig.~\ref{Neutrality}, we find separate self-consistent gapped solutions with and without \CT
symmetry breaking. The \CT symmetry broken solution features moir\'{e} bands
with well-defined non-zero Berry curvatures and sublattice polarizations $\rho_B-\rho_A$, and
 is lowest in energy in most regions of the phase diagram as shown in Fig.~\ref{Summary}(a).

In Table~\ref{Table1} we note that the ratio of the cost in intra-layer tunneling energy,
to the energy gain from inter-layer tunneling in the non-interacting ground state is $1:2$, the ratio
that is obtained when inter-layer tunneling is treated as a weak perturbation.  This observation is 
consistent with the property \cite{BMModel} that the first magic angle in twisted bilayer graphene is accurately 
predicted by perturbation theory.   Secondly we observe that at neutrality both non-interacting and interacting 
ground states have almost uniform charge density, even though the flat band wavefunctions are
spatially peaked near AA positions in the moir\'e pattern.  The absence of a Hartree energy 
at neutrality is related to the moir\'e band Hamiltonian's approximate 
particle-hole symmetry, and is quite distinct from what would be obtained 
if the Hilbert space were truncated to include only the lowest energy flat bands.
Finally we note that the condensation energy of the gapped state, which 
we define as the difference between ground state energy and the expectation 
value of the Hamiltonian in the non-interacting ground state, is 
$\sim 117$ meV per moir\'e period
and originates  mainly from enhanced interlayer exchange energies.  
At this value of $\epsilon^{-1}$ and $\theta$, total energy minimization including interactions adjusts the ground state 
so as to enhance interlayer tunneling and interlayer exchange energies at a cost in the intra-layer hoping energy,
and a substantial part of the ground state rearrangement occurs in higher energy (remote) valence bands.
A description of interaction physics in terms of the single-particle flat bands alone is sufficient only for $\epsilon^{-1} \lesssim 0.04$.
Even in this case, however, it is necessary to include the Hartree \cite{GuineaInteractions} and exchange self-energies from the frozen negative energy sea which 
lower the energies of states near the moir\'e BZ (mBZ) $\kappa$ and $\kappa'$ points (see Fig.~\ref{Neutrality}) relative to those near $\gamma$ and 
therefore contributes to quasiparticle band dispersion.

\begin{figure*}[t]
	\begin{center}
		\centering
		\includegraphics[width=1.5\columnwidth]{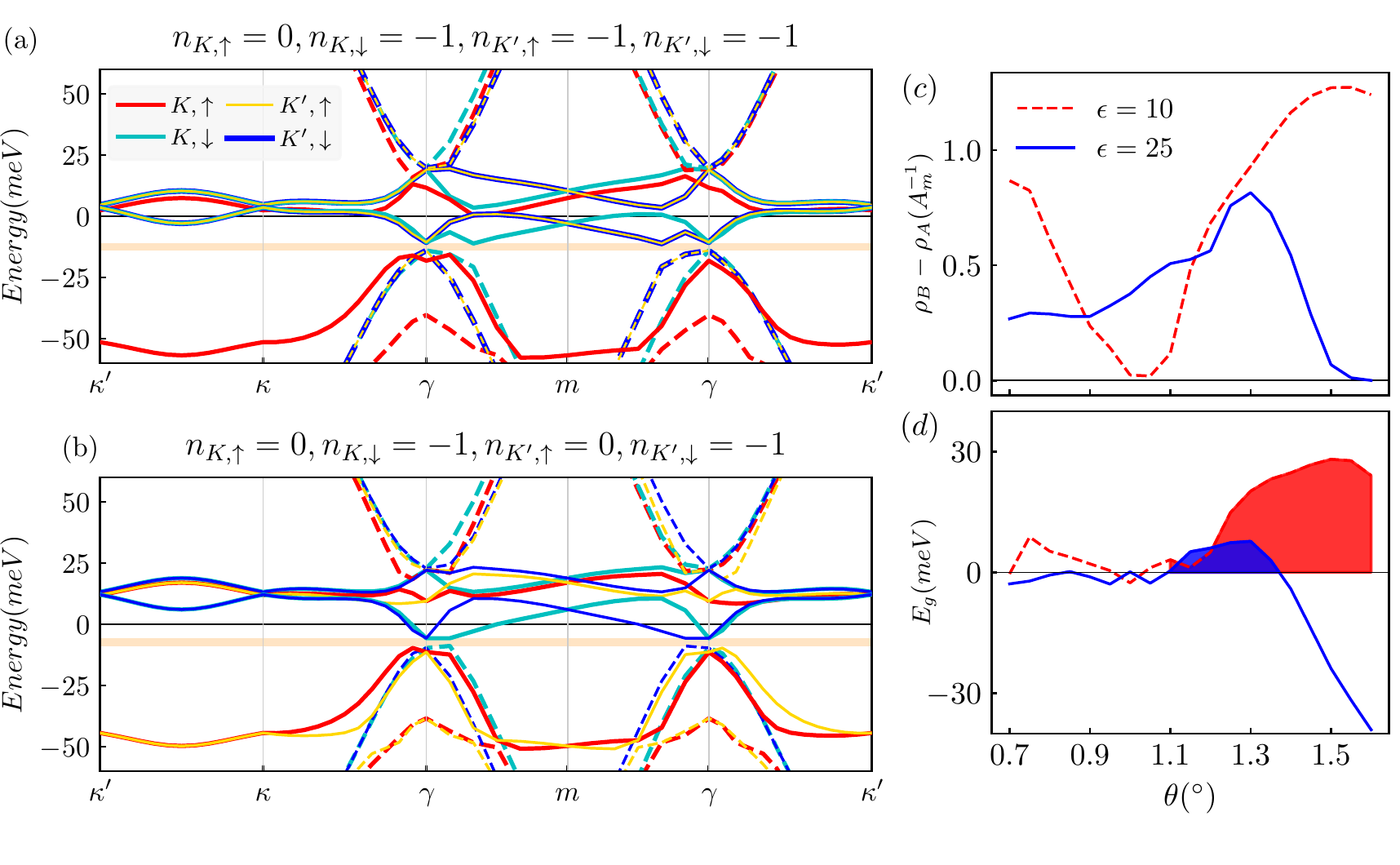}
		\caption{Quasiparticle dispersion of the SCHF ground states for $\theta=1.1^{\circ}$ and $\epsilon=10$ at filling factors:
			(a) $n=-3/4$ and  (b) $n=-1/2$.
			The lowest conduction bands and the highest valence bands are plotted as solid lines 
			while remote bands are plotted as dashed lines with flavor-dependent colors.
			The charge gaps (shaded gold) are $E_g=2.86$ meV for $n=-3/4$  and $E_g=3.15$ meV for $n=-1/2$.
			The flavor dependent band occupation numbers $n_{K/K',\uparrow/\downarrow}$ are measured from neutrality so that 
			$0$ means that the corresponding valence band is occupied and $-1$ means that it is 
			empty. 
			(c)  \CT order parameter and (d) global energy gap as a function of twist angle at $n=-1/2$ in a state with 
			one flat valence band occupied for each valley. 
			The red (dashed) and blue (solid) lines correspond to $\epsilon=10$ and $25$, respectively.
			The flavor polarized insulators are metastable, i.e. the assumed gap is self-consistent,
			in an interaction strength dependent interval (shaded regions) 
			 on the high-twist angle side of the magic angle.}
		\label{MultipleFlavorDispersion}
	\end{center}
\end{figure*}

{\em Band topology, insulating states, and broken \CT symmetry.}---\noindent
The moir\'e band model \cite{BMModel} captures the 
microscopic tight-binding model's $\mathcal{D}_6$,  
time reversal $\mathcal{T}$, and $U(1)$ valley symmetries. 
When the Hamiltonian is projected to a single valley, 
both $\mathcal{T}$ and $\mathcal{C}_2=(\mathcal{C}_6)^3$a are lost because they map states between valleys.
We are left only with the combined symmetry \CTs,
the three-fold rotational symmetry $\mathcal{C}_3$ and a
two-fold rotation with respect to the $x$-axis $\mathcal{M}_x$.
As in monolayer graphene, the $\mathcal{C}_3$ symmetry guarantees 
Dirac points at both $\bm{\kappa}$ and $\bm{\kappa'}$
in the mBZ.  We find that near magic angle $\mathcal{C}_3$ is already broken
at the weakest interactions we consider and that
because of the flatness of the magic-angle bands, the Dirac point positions rapidly move close to
$\gamma$ where the bands are most dispersive. 
The band topology evolution with $\theta$ and $\epsilon^{-1}$ is 
sensitive not only to interactions, but also to the details of the non-interacting band model (see Supplemental Material and \cite{XieMacDonald}).
Unless otherwise specified, we have taken $T_{AA}/T_{AB}=0.8$, where $T_{AA}$ and $T_{AB}$ are the continuum model's 
intra-sublattice and inter-sublattice hopping parameters, to account
for corrugation and strain effects \cite{Kaxiras}.
Even though this choice gives 
a gap between flat bands and remote bands in the non-interacting 
limit, near the magic angle a small interaction strength is sufficient to pull
the highest valence (lowest conduction) band down (up)
in energy to touch the remote bands. 
The band topology at $\epsilon=5$ is illustrated in 
Fig.~\ref{Neutrality}(b) by plotting Berry curvature in the 
\CT broken state as a function of moir\'e band momentum.
Because of the involvement of remote bands, \CT symmetry no longer
guarantees degeneracies between the first conduction and valence bands
\cite{VishwanathWannierObstructions,VishwanathMarch}.
Breaking \CT symmetry does however lift degeneracies between flat and remote bands,
and generates corresponding Berry curvature hot spots that are visible in Fig.~\ref{Neutrality}(b).
The difference in condensation energy between \CT breaking and preserving states is extremely small.
It follows that near magic twist angles \CT symmetry breaking is 
not essential for gap formation at moderate and stronger interaction strengths. 

We characterize states that do break \CT symmetry
by performing a Pauli matrix expansion of mBZ averages 
of the non-local Fock exchange self-energy, defining
\begin{equation}
\frac{A_{m}}{A} \sum_{\vec{k}}^{mBZ} \langle \bm{k}+\bm{G}, l',s' |\Sigma^{F} | \bm{k}+\bm{G}, l,s\rangle = 
\sum_{ij} \Delta_{\bm{G},ij}  \, \sigma^i_{s's} \tau^j_{l'l}. 
\label{c2t}
\end{equation}
where $A$ is the system area, $A_{m}$ is the moir\'e  unit cell area,
$(i,j)=0,\ldots,3$ are Pauli matrix labels, $(s's)=A,B$ are sublattice 
labels and $(l'l)=t,b$ (top, bottom) are layer labels.  
(The role of self-energy terms that are off-diagonal in reciprocal lattice vector 
as \CTs-breaking order parameters
is discussed in the Supplemental Material.)
The slow fall-off of the order parameter's reciprocal lattice vector expansion reflects the spatial scale of 
quasiparticle wavefunction variation within the moir\'e unit cell.  The largest symmetry-breaking self-energies
are proportional to $\sigma^z_{s's}$ and $\tau^0_{l'l}$, 
{\it i.e.} they are layer independent mass terms that favor one sublattice over the other.
(We have sought self-consistent solutions with large $\sigma^z_{s's} \tau^z_{l'l}$, self-energies 
but find that they are not stable.)  
The Berry curvatures plotted in Fig.\ref{Neutrality}(b)
are large near $\bm{\gamma}$, not near $\bm{\kappa}, \bm{\kappa'}$, as they
would be if the same self-energy were added to a weakly-coupled-layer 
band Hamiltonian at a larger twist angle.
We find that the flat bands sometimes have non-zero Chern numbers, implying 
that quantized anomalous Hall effects can occur \cite{ZhangChern} when band occupations 
are valley-dependent.  

{\em Flavor symmetry breaking.}--- Because interactions normally induce gaps between conduction and 
valence bands, which then remain relatively flat, states with spin/valley flavor dependent band occupancies can be 
insulating. For example at $n=+p/4$, a state with the first conduction band occupied for $p$ flavors and empty for 
the remaining flavors is metastable if the exchange-energy shift of the conduction band upon
 occupation $U_X$ exceeds the conduction band width.  A rough estimate based on non-self-consistent 
 Hartree-Fock calculations (see Supplemental Material) yields $U_X \simeq 250 {\rm meV}/\epsilon$,
 increasing slowly with twist angle.  The flat band width, on the other hand, increases rapidly when the 
 magic angle is exceeded, so that insulating states are restricted to the 
immediate vicinity of the magic angle.  Fig.~\ref{MultipleFlavorDispersion} illustrates the 
quasiparticle bands that emerge from a typical fully self-consistent 
mean-field calculation for a broken flavor symmetry insulator.  
In mean-field theory, coupling between flavors occurs
only through the Hartree potential which is absent at neutrality
and attractive at $AA$ sites in the moir\'e pattern at negative band filling factors.  It follows that 
the energies and wave functions of the Hartree-Fock quasiparticle states of one flavor depend 
only weakly on the band-fillings of the other flavors.

{\em Discussion.}--- Guided by earlier work \cite{Jung2014}, we anticipate that there is generally an energetic 
preference (not captured in continuum models) for states in which opposite valleys are occupied equally. 
insulating states  at $n=\pm 1/4$ and $\pm 3/4$, must however break valley symmetry and are likely to be maximally
valley polarized, which implies quantization of the 
anomalous Hall effect.  It therefore follows from our calculations that quantized anomalous Hall 
effects in graphene bilayers can occur at quarter band fillings, 
but that it may not either because \CT symmetry is not broken or because the 
Chern number happens to equal zero.  Indeed there is evidence \cite{Efetov} experimentally
that some insulating states at the quarters are Chern insulators and some are not.  
Further experimental work that maps out how this behavior depends on twist angles 
and distances to gates will be necessary to make a detailed comparison with mean-filed theory.
The important role of thermal and quantum fluctuations of the collective fields present in the insulating states 
will be discussed elsewhere.  

\noindent
\begin{acknowledgements}
{\em Acknowledgment.}---\noindent
We acknowledge support from DOE BES under Award DE-FG02-02ER45958 and helpful
interactions with A. Bernevig, D. Efetov, P. Potasz, N. Regnault, T. Senthil, A. Vishwanath, and F. Wu.  
\end{acknowledgements}

{\em Note added.}---\noindent
Three independent related papers\cite{Ashvin19, ValleyWave, Dai19} which appeared 
after submission of our work report related results from mean-field theory and are complimentary to this paper.

\newpage
\appendix

\section{Supplemental Material}
\renewcommand\thefigure{S\arabic{figure}}
\setcounter{figure}{0}
\renewcommand\theequation{S\arabic{equation}}
\setcounter{equation}{0}

\subsection{Self-consistent Hartree-Fock theory \\ for twisted bilayer graphene}
Our mean field calculation is based on the continuum model proposed in \cite{BMModel}.
Below we first explain the formalism for the flavorless case, {\it i.e.} the valley projected and spinless case,
and then extend it to include both spin and valley degrees of freedom.
We choose the convention that the top and the bottom layers are respectively rotated counter-clockwise and 
clockwise around the $\hat{z}$ axis by a small angle $\theta/2$.
The relative displacement between layers prior to rotation does not affect the spectrum and is ignored in our tunneling 
Hamiltonian.  The quasiparticle wave functions in this theory are four component spinors, 
$\psi_{\alpha, \bm{k}}(\bm{r})$, $\alpha=\{A1, \ B1,\ A2,\ B2\}$, where $A(B)$ in the first index specifies sublattice 
and $1(2)$ in the second index specifies layer.  
We solve the mean-field equations by expanding each component in a plane-wave 
basis consistent with the continuum model's spatial periodicity, which matches that of the moir\'e pattern.  

The spinless single particle Hamiltonian projected onto valley K takes the form:
\begin{align}
\mathcal{\hat{H}}^{\rm{K}}_{0} = 
\begin{pmatrix}
h_{\theta/2}(\bm{k}) & h_{T}(\bm{r})\\
h^\dagger_{T}(\bm{r})  & h_{-\theta/2}(\bm{k}')
\end{pmatrix} \label{ham}
\end{align}
where $\hat{h}_{\pm \theta/2}$ are the Dirac Hamiltonians for isolated rotated graphene layers,
\begin{align}
h_{\theta}(\bm{k}) = -\hbar v_D |\bar{\bm{k}}| 
\begin{pmatrix}
0 & e^{i (\theta_{\bar{\bm{k}}}- \theta)} \\
e^{-i  (\theta_{\bar{\bm{k}}}- \theta)}  & 0
\end{pmatrix},
\end{align}
$\theta_{\bar{\bm{k}}}$ is the orientation angle of momentum measured from the Dirac point 
$\bar{\bm{k}}=\bm{k}-\bm{K}_{\theta}$.  ($\bm{K}_{\pm\theta/2}$ is the Dirac momentum of top(bottom) layer.)
Due to the moir\'{e} periodic modulation, interlayer tunneling is accompanied by momentum boosts 
$\bm{q}_j=\bm{0}$, $\bm{b}_1$ or $\bm{b}_2$ for $j=0, 1, 2$ respectively.
$\bm{b}_{1,2}=(\pm 1/2,\sqrt{3}/2)4\pi/(\sqrt{3}a_M)$
are the basis vectors of moir\'{e} reciprocal lattice, where $a_M=a/(2\sin(\theta/2))$ is the lattice constant of moire pattern 
and $a$ the lattice constant of monolayer graphene.
The spatial modulation of the tunneling Hamiltonian takes the form:
\begin{align}
h_T(\bm{r}) = \sum_{j=0}^3 T_j e^{-i\bm{q}_j\cdot \bm{r}}
\end{align}
where
\begin{align}
T_j = \omega_0\sigma_0 + \omega_1\cos(j\phi)\sigma_x + \omega_1\sin(j\phi)\sigma_y
\end{align}
Note that we have chosen a different convention for $T_j$ than in Ref.~\onlinecite{BMModel}
which sets the origin of real space at the center of the AA region. 
The magnitude of interlayer tunneling is taken to be $\omega_1\equiv T_{AB}=110meV$
and $\omega_0\equiv T_{AA}=0.8\omega_1$.
Below $\bm{k}$ is understood to be restricted
to the first moir\'{e} Brillouin zone (mBZ) ($\bm{k} \in$ mBZ).
For each  $\bm{k}$ we employ the plane-wave expansion basis $|\psi_{\alpha, \bm{G},\bm{k}}\rangle$
where $\bm{G}=m\bm{b}_1+n\bm{b}_2$ and $m,n$ are integers.
The single-particle Hamiltonian $\mathcal{\hat{H}}^{\rm{K}}_{0}$ has both terms
that are diagonal in reciprocal lattice vector and terms that are 
off-diagonal in reciprocal lattice vector.  

We take electron-electron interactions into account using the self-consistent Hartree-Fock method.
In a plane wave basis,  the Hartree and Fock self-energies are
\begin{align}
\Sigma^{H}_{\alpha, \bm{G}; \beta, \bm{G}'}(\bm{k}) 
= \frac{1}{A} \sum_{\alpha'}
V_{\alpha'\alpha}(\bm{G}'-\bm{G}) 
\delta \rho_{\alpha'\alpha'}(\bm{G}-\bm{G}')  \delta_{\alpha\beta}
\label{hartreese}
\end{align}
and
\begin{align}
\Sigma^{F}_{\alpha, \bm{G}; \beta, \bm{G}'}(\bm{k}) 
=-\frac{1}{A}\sum_{\bm{G}'', \bm{k}'}&
V_{\alpha\beta}(\bm{G}''+\bm{k}'-\bm{k}) \notag\\
&\times\delta \rho_{\alpha, \bm{G}+\bm{G}'';\beta, \bm{G}'+\bm{G}''} (\bm{k}')
\label{fockse}
\end{align}
where $\delta\rho = \rho -\rho_{\rm{iso}}$ is the density matrix defined relative to that
of isolated rotated graphene layers each filled up to the charge neutrality point.
The density-matrix $\rho$ is defined in a plane wave basis as:
\begin{align}
\rho_{\alpha,\bm{G};\beta,\bm{G}'}(\bm{k}) = \sum_{n} \; z^{n*}_{\beta,\bm{G}',\bm{k}}z^n_{\alpha,\bm{G},\bm{k}}
\end{align}
where the summation is over filled bands. We have also used
$\delta \rho_{\alpha\beta}(\bm{G})\equiv\sum_{\bm{k},\bm{G}_1}\delta \rho_{\alpha,\bm{G}_1+\bm{G}; \beta, \bm{G}_1}(\bm{k})$
in Eq.~\ref{hartreese}.
$z_{\alpha, \bm{G},\bm{k}}^n$ is a numerical 
eigenvector in the  plane-wave expansion,
\begin{align}
|\psi_{n,\bm{k}} \rangle = \sum_{\alpha, \bm{G}} z_{\alpha, \bm{G},\bm{k}}^n |\psi_{\alpha, \bm{G}+\bm{k}}\rangle,
\end{align}
obtained by diagonalizing the total Hamiltonian
\begin{align}
\mathcal{H} = \mathcal{H}^{\rm{K}}_0 +  \Sigma^{\rm{H}} + \Sigma^{\rm{F}}.
\end{align}

Although interactions alter the lowest conduction and valence bands most strongly,
we find it is generally essential to include a relatively large number of remote conduction and valence 
bands.  The numerical results presented in the MS used a reciprocal space cut-off that allowed a total of 148 bands.
This reciprocal space cut-off is sufficient to achieve converged insulating state 
condensation energies for the range of $\epsilon$ values we have considered.  
Spin and valley flavor degree of freedom are easily restored simply 
by observing that the Hartree self-energy has contributions from all occupied quasiparticle
states of all flavors whereas 
the exchange self-energy  has contributions from occupied quasiparticles 
of the same flavor only.  We note that the Fock self-energy is strongly non-local and that 
this non-locality plays an essential role in how interactions select the character 
of the insulating states.  No local exchange-correlation approximation can 
describe the interaction physics of twisted bilayers reliably.  

\subsection{Decomposition of the Fock self-energy and symmetry analysis}
The Fock self-energy dominates the interaction physics in the SCHF Hamiltonian
because it is responsible for
the broken symmetries of TBG.  In the following we discuss the expansion of 
the Fock self-energy in terms of layer and sublattice Pauli matrices.
For each flavor we can classify terms in this expansion 
according to whether or not they violate the \CT symmetry.
\CT symmetry is a property of the non-interacting Hamiltonian and 
must be broken in order to generate quasiparticle bands with non-zero momentum-space 
Berry curvatures and Chern numbers.  

We can quite generally write:
\begin{equation}
\frac{A_m}{A} \sum_{\vec{k}}^{BZ} \langle \bm{k}+\bm{G}', l,'s' |\Sigma^{F} | \bm{k}+\bm{G}, l,s\rangle = 
\sum_{ij} \Delta_{\bm{G}'\bm{G},ij}  \, \sigma^i_{s's} \tau^j_{l'l}. 
\end{equation}
For $\bm{G}'=\bm{G}$, $\Delta_{\bm{G},ij}\equiv\Delta_{\bm{G}\bm{G},ij}$ is always real as discussed in the main text.
For $\bm{G}'\neq \bm{G}$, $\Delta_{\bm{G}'\bm{G},ij}$ is complex and,
since $\Sigma^{F} $ is Hermitian, must satisfy
$\Delta_{\bm{G}'\bm{G},ij}=\Delta^*_{\bm{G}\bm{G}',ij}$.
If we write 
\begin{align}
\Delta_{\bm{G}'\bm{G},ij}=\Delta^R_{\bm{G}'\bm{G},ij}+i\Delta^I_{\bm{G}'\bm{G},ij},
\end{align}
the $\Delta^R_{\bm{G}'\bm{G},ij}$ terms have the same symmetry classification under 
\CT as the $\Delta_{\bm{G}\bm{G},ij}$ terms, 
i.e., states with finite $\Delta^R_{\bm{G}'\bm{G},ij}$ where $(ij)\in \mathcal{S}\equiv\{(02),(12),(22),(30),(31), (33)\}$ break \CT symmetry.  
On the other hand, the $\Delta^I_{\bm{G}'\bm{G},ij}$ terms have the opposite classification because 
of the complex conjugation operation in the time reversal operator
such that terms $\Delta^I_{\bm{G}'\bm{G},ij}$ with $(ij)$ outside $\mathcal{S}$
break \CT symmetry.

\begin{figure}[t!]
	\begin{center}
		\includegraphics[width=0.99\columnwidth]{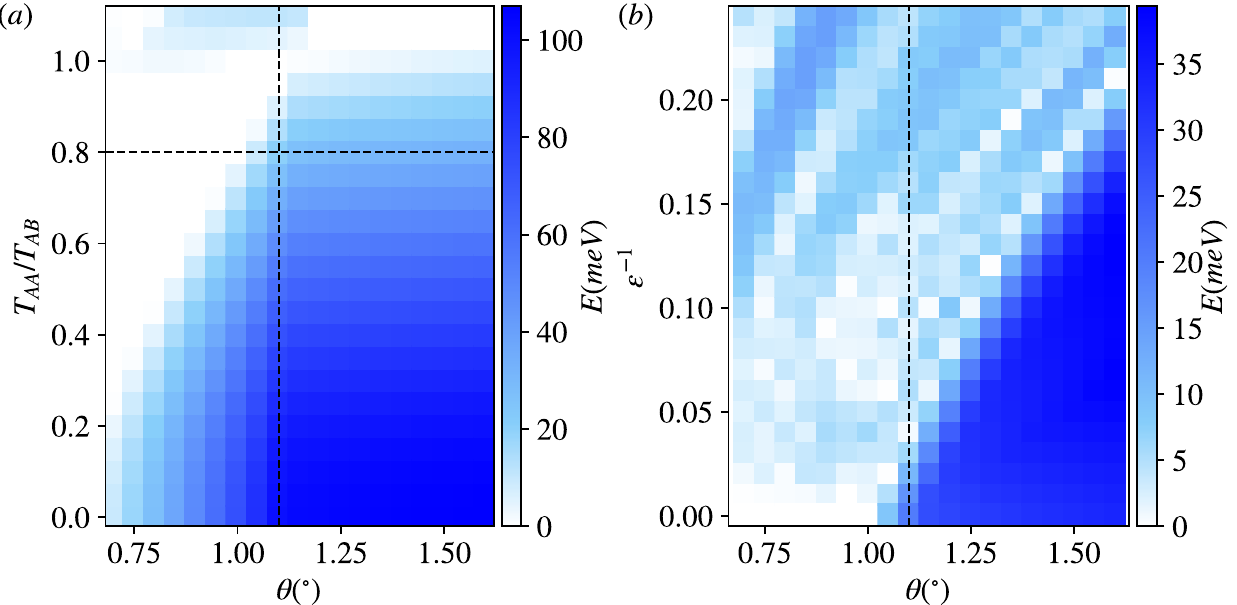}
		\caption{Energy separation between the highest valence band and remote bands in the 
			(a) non-interacting model as a function of twist angle $\theta$ and 
			the ratio of intra-/inter-sublattice tunneling amplitudes $T_{AA}/T_{AB}$ and 
		   (b) SCHF ground states as a function of twist angle $\theta$ and the interaction strength 
			$\epsilon^{-1}$ at $T_{AA}/T_{AB}=0.8$.  Crossings between flat and remote bands occur 
			within the narrow gap (white) regions in (a).  
			Vertical and horizontal dashed lines mark $\theta=1.1$ and $T_{AA}/T_{AB}=0.8$, respectively.  } 
		\label{gapmap}
	\end{center}
\end{figure}

\begin{figure}[t!]
	\begin{center}
		\includegraphics[width=\columnwidth]{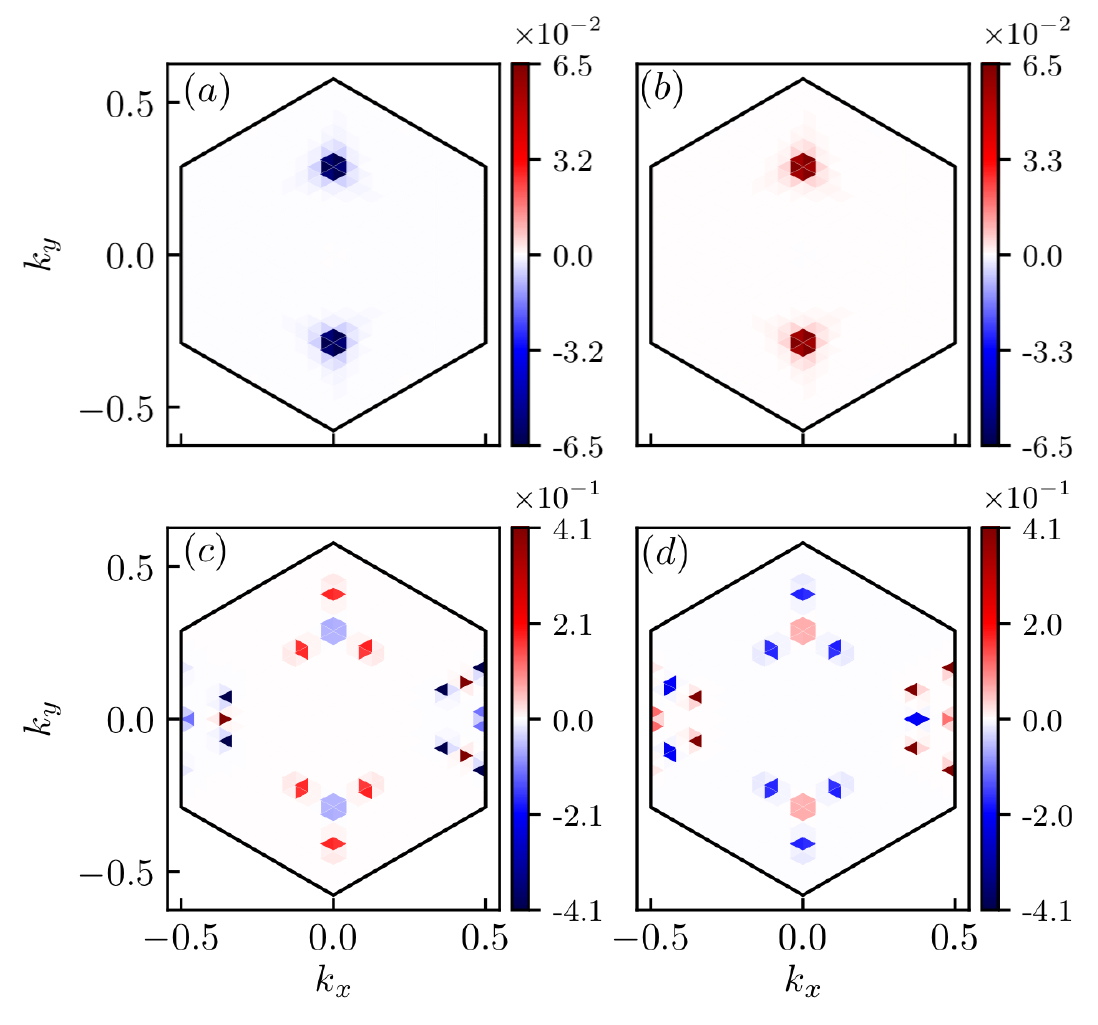}
		\caption{Berry curvature of non-interacting flat bands. Flat bands are isolated from remote band for $\theta=1.1^{\circ}$ (top row)
			and have band crossings with remote bands for $\theta=0.8^{\circ}$ (bottom row).
			The additional Berry curvature in the later case originates from the band touching with remote band near the $\bm{\gamma}$ point.
			(a) and (c) are for flat conduction bands while (b) and (d) are for flat valence bands.
			A small positive mass term proportional to $\sigma_3 \tau_0$ has been added to the Hamiltonian to
			break \CT symmetry and gap the degeneracy points in order to reveal the Berry curvature.
			We use $T_{AA}/T_{AB}=0.8$ as in the main text.} 
		\label{SPberry}
	\end{center}
\end{figure}

\subsection{Band topology evolution with twist angle and interaction strength}
Fig.~\ref{gapmap}a shows the non-interacting gap between the flat band and remote bands 
which depends on both twist angle $\theta$ and on the ratio $T_{AA}/T_{AB}$.
At magic angle $\theta=1.1^{\circ}$ (marked by the vertical black dashed line), 
flat bands are in touch with remote bands in the original BM continuum model $T_{AA}/T_{AB}=1$.
To approximate the effect of strain in interacting electron calculations we choose $T_{AA}/T_{AB}=0.8$ unless
otherwise stated.  In Fig.~\ref{gapmap}b we shows the energy separation between the highest 
interaction-modified quasiparticle valence band and the remote valence bands as a function
of twist angle $\theta$ and the interaction strength $\epsilon^{-1}$.

In Fig.~\ref{SPberry} we plot the momentum-space Berry curvature of the non-interacting continuum model 
bands with a small $\sigma_3 \tau_0$ mass term added to break \CT symmetry and 
reveal the Dirac points.  Near the magic angle the band topology is altered relative to the band topology 
of weakly coupled layers at larger twist angles by transferal of Dirac points from remote bands to the 
lowest conduction and valence bands.  The band topology is further altered by electron-electron interactions 
as illustrated for the self-consistent mean field ground state discussed in the main text (Fig.~2(b)). 
For a twist angle just above the magic angle Fig.~\ref{SPberry} shows that the weak-coupling $\pm \pi $ Berry phases near 
$\kappa$ and $\kappa'$ are surrounded by three new Dirac points with the opposite chirality bands.  The change 
in topology occurs  at a slightly higher twist angle than illustrated here and is accompanied by 
band touchings involving remote bands that occur at momenta near $\gamma$. 
A detailed discussion of the dependence of non-interacting band topology on twist angle and on details of the 
interlayer hopping model will be presented elsewhere \cite{XieMacDonald}.

\subsection{Exchange-energy shift and nature of quasiparticle bandgap}

Exchange interaction shifts the quasiparticle bands of a particular flavor downward in energy 
as the electron filling of this flavor increases.
The quasiparticle bandgap of the flavor symmetry breaking states thus forms  
between the same active bands of different flavors, provided that the renormalized bandwidth
is less than the amount of exchange-energy shift.
Here we estimate the exchange-energy shift in a non-self-consistent (NSC) manner. 
As an example, we examine the case $p=1$, i.e., 
with the lowest conduction band of one particular flavor filled and the rest of the flavors filled to neutrality.

\begin{figure}[t!]
	\begin{center}
		\centering
		\includegraphics[width=0.99\columnwidth]{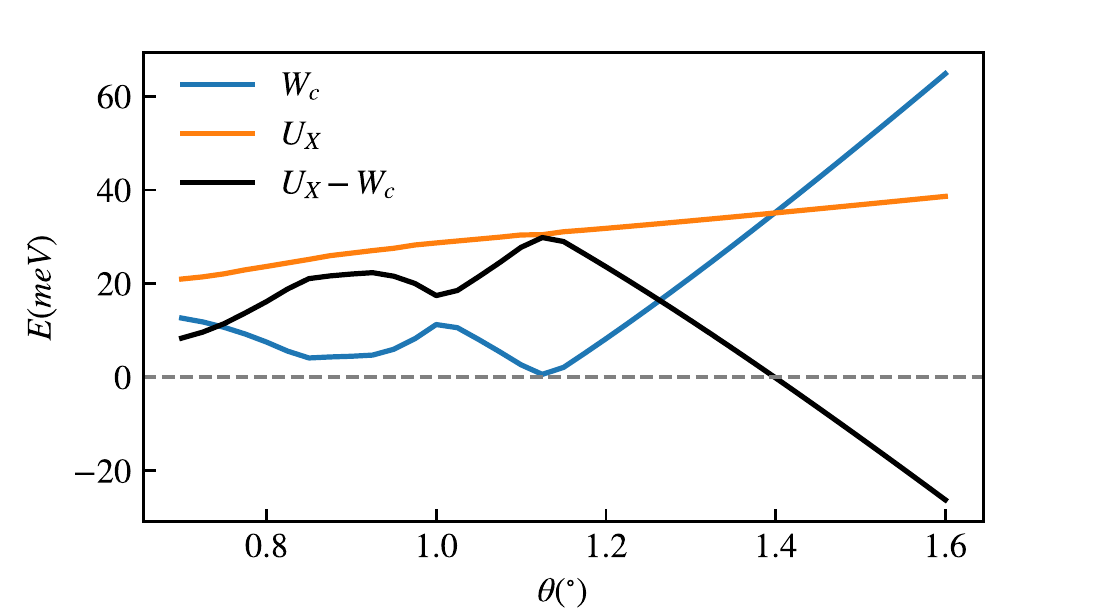}
		\caption{Exchange-energy shift $U_X$ and renormalized bandwidth $W_c$ 
			of the filled conduction band for $n=1/4$ as a function of twist angle. 
			The interaction strength is fixed at $\epsilon=10$. 
			In order to have broken symmetry insulating solutions that are at least metastable, it is required that 
			$U_X > W_c$.}
		\label{exchangeshift}
	\end{center}
\end{figure}

Define the amount of energy lowered by exchange self-energy from the filled flat conduction band as
\begin{align}
U_X = - \frac{A_m}{A}\sum_{\bm{k}}\langle \psi^{\textrm{NSC}}_{c,\bm{k}}  |\Sigma^F(\delta\rho^{\textrm{NSC}}_c)| \psi^{\textrm{NSC}}_{c,\bm{k}}\rangle,
\end{align}
where $\psi^{\textrm{NSC}}_{c,\bm{k}}$ is the lowest conduction band wavefunction considering the self-energy of the frozen remote bands
and $\delta\rho^{\textrm{NSC}}_c$ the corresponding density matrix.
$\Sigma^F$ is the Fock self-energy defined in Eq.~\ref{fockse}.
Fig.~\ref{exchangeshift} compares $U_X$ and the renormalized bandwidth $W_c$ of the filled conduction band as twist angle is varied.
It captures the overall trend of the self-consistent value for $U_X-W_c$ that peaks near to but above the magic angle.   

\begin{figure}[t!]
	\begin{center}
		\centering
		\includegraphics[width=0.99\columnwidth]{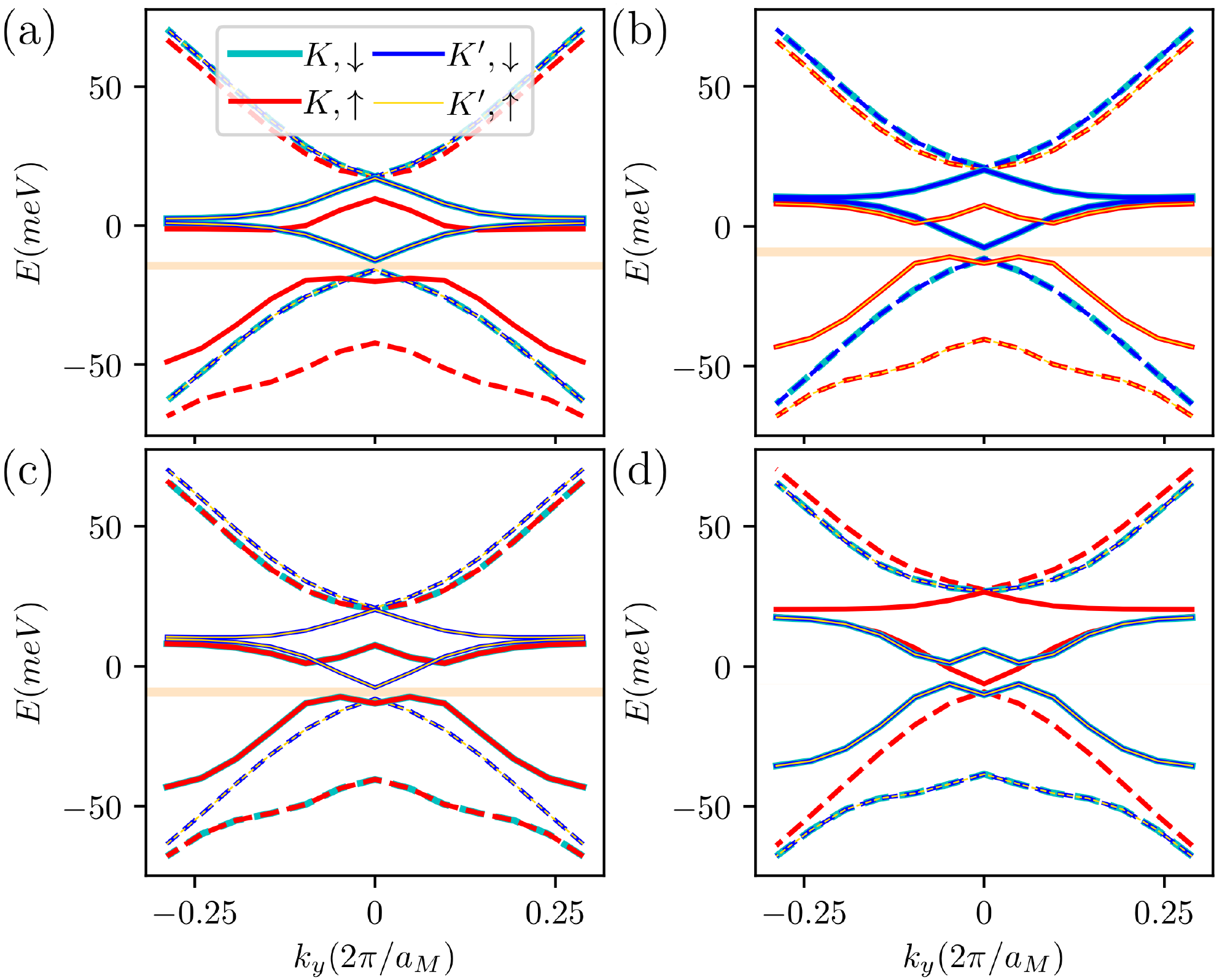}
		\caption{Quasiparticle dispersion near bandgap for flavor symmetry breaking states
			(a) $n=-3/4$, (b) $n=-1/2$ with spin polarization, (c) $n=-1/2$ with valley polarization
			and (d) $n=-1/4$.  The dispersions are drawn along the line which passes $\gamma$ 
			and is parallel to $k_y$-direction.}
		\label{quarter_gap}
	\end{center}
\end{figure}

The exact value of the insulating gap depends on the details of the quasiparticle dispersion and must 
be determined by a converged self-consistent calculation.  
Fig.~\ref{quarter_gap} shows quasiparticle band dispersions near band extrema
 for different flavor symmetry breaking states.
For $n=-1/4$ and $-1/2$, the bandgap occurs between flat valence bands of filled and empty flavors 
and is indirect in momentum.
However, for $n=-3/4$, because the remote valence band is slightly higher than the filled flat valence band,
the bandgap becomes direct.

\begin{figure}[t!]
	\begin{center}
		\centering
		\includegraphics[width=0.99\columnwidth]{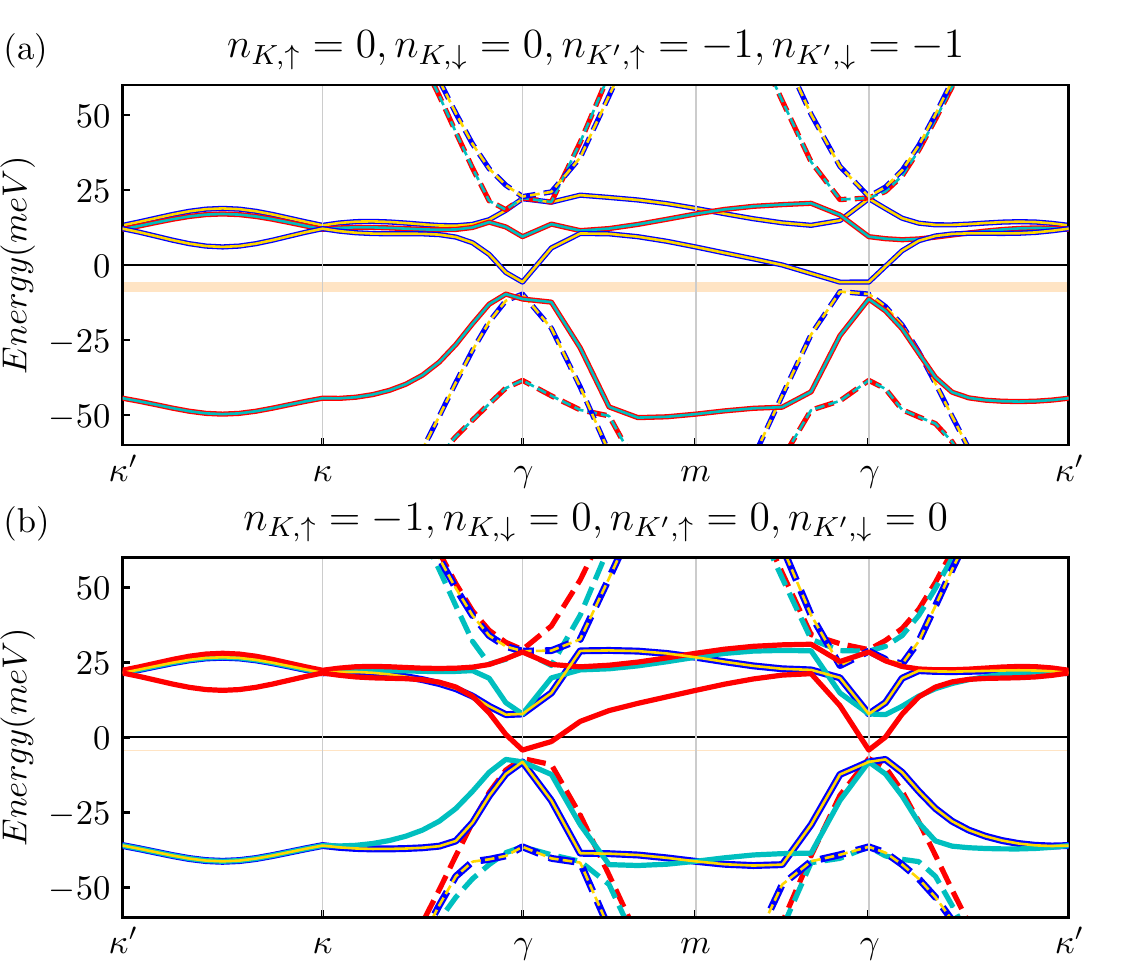}
		\caption{Quasiparticle dispersion of the SCHF ground states at  (a) $n=-1/4$ and  (b) $n=-1/2$ with valley polarization.
			Color scheme and notations are the same as in Fig. 3(a)-(b) of the main text.}
		\label{multiflavor_more}
	\end{center}
\end{figure}

\subsection{Quasiparticle dispersion of $n=-1/4$ and valley polarized $n=-1/2$ states}  %
In this section, we present quasiparticle dispersions for SCHF ground states at filling factors 
$n=-1/4$ and $n=-1/2$ with valley polarization.
Fig.~\ref{multiflavor_more} shows the overall quasiparticle dispersion for the two cases where
the same set of parameters are used as in Fig. 3 (a) and (b).
Both states are insulating with energy gap $E_g=3.16$ meV in the valley polarized $n=-1/2$ state
and $E_g=0.06$ meV in the $n=-1/4$ state.

At $n=-1/2$, mean field states with full filling of either two of the four flavors,
e.g., the valley polarized and the spin polarized states,  are metastable and have degenerate total energies.
We expect that including valley-anisotropic terms, such as short range Coulomb interactions and electron-phonon interactions,
will lift the degeneracy and lead to ground states with inter-valley coherence.
A detailed study of these interaction terms will be included in our future work.

\subsection{Total energy difference \CT preserving and breaking states at neutrality}

\begin{figure}[t!]
	\begin{center}
		\centering
		\includegraphics[width=0.8\columnwidth]{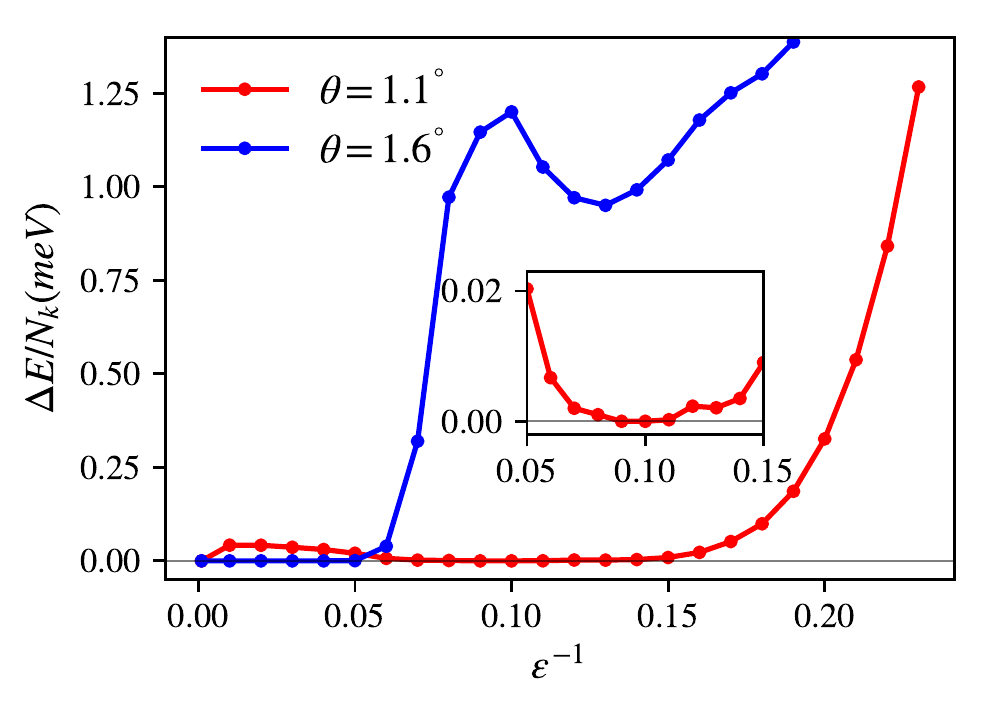}
		\caption{Difference in total energy per moir\'{e} unit cell per flavor between the \CT preserving solution and the lowest energy ground state at neutrality.
			The inset is a zoom in near $\epsilon^{-1}=0.1$ for $\theta=1.1^{\circ}$ where the \CT preserving state is the ground state, i.e. $\Delta E =0$.
		    For $\theta=1.1^{\circ}$, a tiny range of interaction strength near zero interaction strength exists that do not break \CT symmetry which was not captured
	        by the line interpolation between our discrete sampling points.}
		\label{totalenergy}
	\end{center}
\end{figure}

The self-consistent Hartree-Fock calculation yields solutions either preserve or break \CT symmetry.
In Fig.~\ref{totalenergy}, we present the total energy difference between the lowest energy \CT preserving solution 
and the lowest energy solution as a function of interaction strength for the magic angle case $\theta=1.1^{\circ}$,
and for the large twist angle case $\theta=1.6^{\circ}$.  
In the range of interaction strength where the energy difference vanishes, we only found \CT preserving solutions.
Near the magic angle, the energy gained tends to be small even when \CT symmetry is broken.
For $\theta=1.6^{\circ}$.
starting from small interaction strength limit, there is a phase transition from a gapless state with \CT symmetry to a gapped \CT breaking state
which occurs at about $\epsilon^{-1}\approx 0.06$.
For magic angle case, this transition occurs first at a very small interaction strength which is not captured by our discrete plot, but does not produce a large
ground state energy reduction.  At twist angle $\theta=1.6^{\circ}$, the cusp of the curve after the transition is believed to be associated with
the complicated band crossing occurring between flat and remote bands.

\subsection{Flat band projected Hartree and Fock self-energy}

\begin{figure*}[t]
	\begin{center}
		\centering
		\includegraphics[width=1.8\columnwidth]{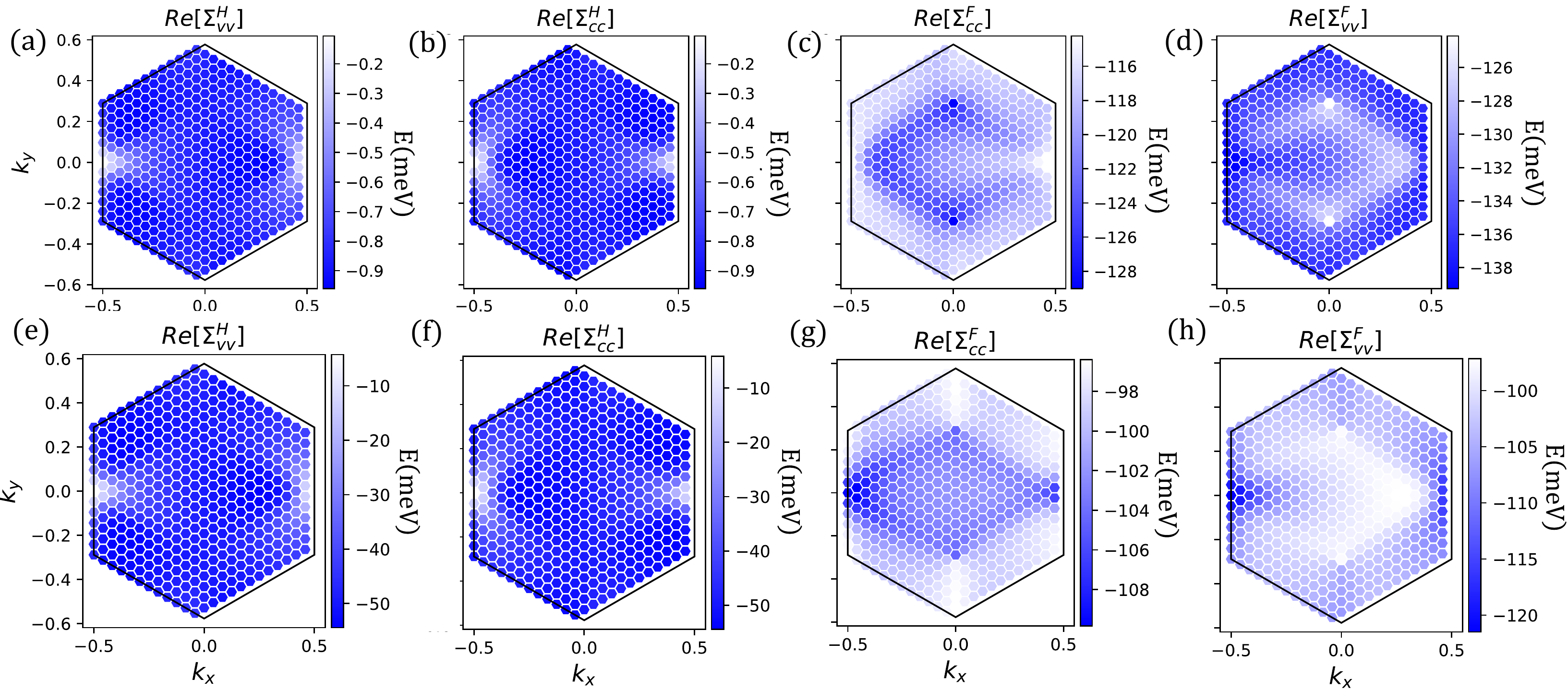}
		\caption{Projection of Hartree and Fock self-energies onto non-interacting flat bands at neutrality (a-d) and at empty flat bands (e-f). 
			$\Sigma^H_{ab}$ and $\Sigma^F_{ab}$ stand for Hartree and Fock self-energies 
			respectively where $(a,b)=(c,v)$ are band indices for flat conduction (c) and valence (v) bands.
			Only the  diagonal components of the projection are plotted and these have vanishing imaginary parts.}
		\label{HFselfenergy}
	\end{center}
\end{figure*}

To have a better understanding of the different roles played by Hartree and Fock interaction self-energies 
on the low energy flat bands, we plot the projection of the self-energies to the Hilbert space of the non-interacting flat bands.
We note this is done in a non-self-consistent fashion in which the self-energy are constructed from non-interacting 
density matrices. 
Fig.~\ref{HFselfenergy} plots the different components of the projected self-energies at neutrality and in the case 
in which the flat bands are all empty. 
At neutrality, the Hartree self-energy is vanishingly small and almost does not affect the flat band dispersion.
However, the Fock self-energy plays an important role in changing the flat band states by lowering the energy of the filled
valence band more strongly relative to the energy of the conduction band.  
In contrast, at empty filling of the flat valence bands, Hartree energy becomes finite.
Because the Hartree self-energy is common to all the flavors, it does not play a role in splitting 
the energy between filled and empty flavors.
The Fock self-energy is diagonal in flavor and differs between filled and empty flavors which result in exchange-energy shift responsible for flavor symmetry
breaking as discussed in previous section.  It does however alter the quasiparticle dispersions by lowering the energies of states that are more strongly
localized near AA-stacking positions in the moir\'e pattern relative to states that are less strongly peaked.

\subsection{Non-interacting eigenstates at the high symmetry momenta  \texorpdfstring{$\bm{\gamma}$}{$\gamma$} and \texorpdfstring{$\bm{m}$}{$m$}}

A minimal non-interacting Hamiltonian at  the $\bm{m}$ point can be constructed by keeping the lowest order terms in 
Eq.~\ref{ham}: 
\begin{align}
	H_0(\bm{m}) = -\frac{1}{2} \hbar v_D k_{\theta} \sigma_y \tau_z + \omega_0 \sigma_0 \tau_x + \omega_1 \sigma_x\tau_x.
	\label{ham_m}
\end{align}
This Hamiltonian acts on the four-component spinor 
$(\psi_{A1, \bm{k}_{\bm{m}}}, \psi_{B1, \bm{k}_{\bm{m}}}, \psi_{A2, \bm{k}_{\bm{m}}}, \psi_{B2, \bm{k}_{\bm{m}}})^T$ 
where $\bm{k}_{\bm{m}}$ is the momentum of $\bm{m}$.
$\mathcal{H}_0^K(\bm{m}) $ can be rewritten in a block diagonalized form by simultaneous rotation around 
the $y$ axis in both sublattice and layer pseudo-spin spaces so that
$\sigma_x\rightarrow\sigma_z$, $\sigma_z\rightarrow-\sigma_x$ ,$\tau_x\rightarrow\tau_z$, and $\tau_z\rightarrow-\tau_x$. 
We then obtain the eigenstate energies,
\begin{align}
	E_{\bm{m}} = \pm \omega_1 \pm \sqrt{\hbar^2 v_D^2 k_{\theta}^2/4+\omega_0^2}.
\end{align}
The two states close in energy to the charge neutrality point have energies $\pm(\omega_1 - \sqrt{\hbar^2 v_D^2 k_{\theta}^2/4+\omega_0^2})$
which suggests that a band inversion can happen at $\bm{m}$ as twist angle $\theta$ is decreased when $\omega_1 < \omega_0$.  

At the high symmetry momentum $\bm{\gamma}$, the lowest order Hamiltonian couples degenerate Dirac states of the isolated layers with momentum 
$\Gamma^t_i=\{\bm{\gamma}, \bm{\gamma}+\bm{b}_1, \bm{\gamma}+\bm{b}_1+\bm{b}_2\}$ from the top layer and $\Gamma^b_i=\{\bm{\gamma}, \bm{\gamma}+\bm{b}_2, \bm{\gamma}+\bm{b}_1+\bm{b}_2\}$ from the bottom layer, where $i=0,1,2$.
States at these momenta are coupled in a cyclic order and the resulting Hamiltonian takes the form,

\begin{widetext}
	\begin{align}
		H_0(\bm{\gamma})=
		\begin{pmatrix}
		h_{-\theta/2}(\Gamma^b_0)  & T_0 &0 & 0&0 & T_{1} \\
		T_0 & h_{\theta/2}(\Gamma^t_0)  & T_{-1} &0 & 0&0 \\
		0 & T_{-1} & h_{-\theta/2}(\Gamma^b_1)  & T_{1} &0 & 0\\
		0 & 0 & T_{1} & h_{\theta/2}(\Gamma^t_2)  & T_{0} &0 \\
		0 & 0 & 0 & T_{0} & h_{-\theta/2}(\Gamma^b_2)  & T_{-1} \\
		T_{1} & 0 & 0 & 0 & T_{-1} & h_{-\theta/2}(\Gamma^t_1)
		\end{pmatrix},
	\end{align}
\end{widetext}
which acts on a twelve-component spinor (two sublattices at each of the six momenta.)
We find that $H_0(\bm{\gamma})$ can be diagonalized by first rotating from the sublattice basis 
to the basis which diagonalize the Dirac Hamiltonian $h_{\pm\theta/2}$ at each of the six momenta.
One can group the positive ($+$) and negative($-$) energy Dirac states and further rotate within each group to their eigen-basis.
The final Hamiltonian can be written as,
\begin{align}
	H_0(\bm{\gamma}) = 	H_0^{+}(\bm{\gamma}) + H_0^{-}(\bm{\gamma}) + H_0^{+-}(\bm{\gamma})+ H_0^{-+}(\bm{\gamma})
\end{align}
where
\begin{widetext}
	\begin{align}
	H_0^{\pm}(\bm{\gamma}) = \hbar v_D |\bm{k}_{\bm{\gamma}}|\cdot
	\begin{pmatrix}
	\pm 1 + 2b & 0 & 0 & 0 & 0 & 0 \\
	0 & \pm 1 - 2b & 0 & 0 & 0 & 0 \\
	0 & 0  & \pm 1 + \sqrt{3a^2+b^2} & 0 & 0 & 0\\
	0 & 0 & 0  & \pm 1 + \sqrt{3a^2+b^2} & 0 & 0 \\
	0 & 0 & 0 & 0  & \pm 1 - \sqrt{3a^2+b^2} & 0 \\
	0 & 0 & 0 & 0 & 0  & \pm 1 - \sqrt{3a^2+b^2} 
	\end{pmatrix}.
 \end{align}
\end{widetext}
The off-diagonal block takes the form
\begin{align}
H_0^{+-}(\bm{\gamma}) =\frac{i}{\sqrt{3}} \hbar v_D |\bm{k}_{\bm{\gamma}}|a \cdot
\begin{pmatrix}
-2 & 0 & 0 & 0 & 0 & 0 \\
0 & 2 & 0 & 0 & 0 & 0 \\
0 & 0 & -1 & 0 & 0 & 0 \\
0 & 0 & 0 & 1 & 0 & 0 \\
0 & 0 & 0 & 0 & -1 & 0 \\
0 & 0 & 0 & 0 & 0 & 1 \\		 	
\end{pmatrix}.
\end{align}
Here $|\bm{k}_{\bm{\gamma}}|=k_{\theta}$ is the distance from either of the two Dirac points to $\bm{\gamma}$.
We have defined $a=\frac{\sqrt{3}\omega_0}{2\hbar v_D |\bm{k}_{\bm{\gamma}}|}$ and $b=\frac{\omega_1}{\hbar v_D |\bm{k}_{\bm{\gamma}}|}$. 
The Hamiltonian can then be diagonalized and has eigenenergies that are 
either singly degenerate with $E=\pm\hbar v_D |\bm{k}_{\bm{\gamma}}|\cdot (\sqrt{1+4a^2/3}\pm2b)$
or doubly degenerate with
$E=\pm\hbar v_D |\bm{k}_{\bm{\gamma}}|\cdot (\sqrt{1+a^2/3}\pm\sqrt{3a^2+b^2})$.
The degeneracy of these energies agrees with dimension of the irreducible representations of the little group at 
$\bm{\gamma}$.

The energies of the flat bands at the $\bm{\gamma}$ point for weakly coupled bilayers are singly degenerate and correspond to 
$\pm\hbar v_D |\bm{k}_{\bm{\gamma}}|\cdot (\sqrt{1+4a^2/3}-2b)$.
For strong interlayer coupling or small twist angles,
the doubly degenerate states become the states that are lowest in energy at $\bm{\gamma}$
and the flat bands will be in contact with the remote bands at $\gamma$ (see Fig.~\ref{gapmap}(a)).  
This delicate evolution of band topology with twist angle and hopping parameter ration 
is partially illustrated in Fig.~\ref{SPberry}.


\begin{thebibliography}{99}
\bibitem{footnote} In the continuum model single-electrons states are 
	described by valley-dependent four-component envelope-function spinors that specify position-dependent 
	projections onto $\pi$ orbitals on each of the bilayer's four sub lattices.  In the full $\pi$-band 
	Hilbert space any dependence on position consistent with the emergent moir\'e superlattice periodicity 
	is allowed.  In practice we expand the periodic part of the Bloch spinors at each momentum in the 
	moir\'e Brillouin-zone in reciprocal lattice vectors, truncating at convergence.
\end{thebibliography}
\end{document}